\documentstyle[preprint,prb,aps]{revtex}

\begin{document}

\draft
 
\title{$^{\bf 3}$He impurity excitation  spectrum in liquid $^{\bf 4}$He} 

\author{A. Fabrocini$^{1)}$ and A.Polls$^{2)}$}
\address{
$^1)$ Istituto Nazionale di Fisica Nucleare, 
Dipartimento di Fisica, Universit\`a di Pisa, 
I-56100 Pisa, Italy \protect\\
$^2)$ 
Departament d'Estructura i Constituents de la Mat\`eria, Universitat de Barcelona,
 E-08028 Barcelona, Spain}

\maketitle

\begin{abstract}
We microscopically evaluate the excitation spectrum of the $^3$He impurity in 
liquid $^4$He at T=0 and compare it with the experimental curve  
at the equilibrium density. The adopted correlated basis perturbative 
scheme includes up to two independent phonon, intermediate correlated 
states and the correlation operator is built up with two- and 
three-body correlation functions. The experimental spectrum is well
described by the theory along all the available momentum range. A 
marked deviation from the simple Landau-Pomeranchuck, quadratic behavior 
is found and the momentum dependent effective mass of the impurity increases 
of $\sim50~\%$ at $q\sim1.7~\AA^{-1}$ respect to its $q=0$ value.
No signature of roton-like structures is found. 
\end{abstract}
 
\pacs{67.55.Lf, 67.60.-g, 67.40.Db, Pisa preprint IFUP-TH 59-97}

%\twocolumn

\narrowtext

Both experimentalists and theoreticians have devoted a great deal of 
efforts to measure and explain the characteristics of one $^3$He impurity 
in atomic liquid $^4$He. From the experimental point of view it 
is well known how the impurity chemical potential, $\mu_3$, behaves with 
the temperature and the pressure \cite{Ebner70}, its effective 
mass, $m_3^*$, and quasiparticle excitation spectrum, 
$e(q)$ \cite{Hilton77,Fak90}. 
In ref.\cite{Fak90} the authors found a sizeable deviation from the 
quadratic Landau--Pomeranchuk (LP) spectrum \cite{Landau48}, 
$e_{LP}(q)=\hbar^2 q^2/2m_3^*$, 
in low concentrations $^3$He-$^4$He mixtures. 
$e(q)$ was parametrized in a modified LP (LPM) form as:
\begin{equation}
e_{LPM}(q)=\frac {\hbar^2 q^2}{2 m_3^*}\,
\frac{1}{ 1\,+\, \gamma\ q^2}\, .
\label{eq:LPM}
\end {equation}
The estimated values of the LPM parameters,  
at $P=1.6$ bar and $x_3\sim0.05$ ($x_3$=$^3$He concentration), 
are $m_3^*\sim 2.3~m_3$ and $\gamma\sim 0.13~\AA^2$. 

Microscopic calculations, done in the framework of the correlated 
basis function (CBF) perturbation theory \cite{Fabrocini86}, have 
been able to give good estimates of $\mu_3$ and $m_3^*$ at the 
T=0 $^4$He equilibrium density, $\rho_{eq}=0.02185~ \AA ^{-3}$. Lately, 
a diffusion Monte Carlo approach has provided similar results 
\cite{Ceperley_DMC}. There are also theoretical indications of a 
deviation of the spectrum from the LP form \cite{Bhatt78,Reatto96}. 
The presence of a possible roton-like structure in $e(q)$ 
near the crossing with the $^4$He phonon-roton spectrum was supposed in 
ref.\cite{Pita73} but not confirmed in refs.\cite{Bhatt78,Reatto96}. 
However, in the last work an excitation spectrum quite higher 
than the experimental one was found.

Here we will employ the CBF machinery of ref.\cite{Fabrocini86} 
(hereafter denoted as I) to compute, in a microscopic 
way, the whole impurity spectrum. 
The CBF basis used in I consisted in correlated $n$-phonon 
states
\begin{equation}
\Psi_{{\bf q};{\bf q}_1..{\bf q}_n}
%|{\bf q};{\bf q}_1..{\bf q}_n\rangle 
 =\rho_3({\bf q}-{\bf q}_1-..-{\bf q}_n) 
 \rho_4({\bf q}_1)..\rho_4({\bf q}_n) \Psi_0
\label{eq:nphonon}
\end{equation}
where $\rho_4({\bf k})=\sum_{i=1,N_4} e^{i{\bf k} \cdot {\bf r}_i}$  
is the $^4$He density fluctuation operator
and $\rho_3({\bf k})=e^{ i{\bf k} \cdot {\bf r}_3}$ describes the excitation 
of the impurity. The basis states were then properly normalized. 
$\Psi_0=\Psi_0(3,N_4)$ is the ground state wave function of $N_4$ $^4$He 
atoms plus one $^3$He impurity in a volume $\Omega$, 
taken in the $N_4,~\Omega \rightarrow\infty$ limit, at constant 
$^4$He density, $\rho_4=N_4/\Omega$.

A realistic choice for $\Psi_0(3,N_4) $ is made by applying an 
extended Jastrow--Feenberg correlation operator \cite{Feenberg} to the 
non interacting g.s. wave function
\begin{equation}
\Psi_0(3,N_4)\,= F_2(3,N_4) \, F_3(3,N_4) \,  \Phi_0(3,N_4)\, .
\label{eq:Psi_0}
\end{equation}
$F_{2,3}$ are N--body correlation 
operators including explicit two-- and  three--body dynamical
correlations. $F_2$ is written as a product of two  
body Jastrow, $^3$He-$^4$He and $^4$He-$^4$He correlation functions, 
\begin{equation}
F_2(3,N_4)=\prod_{i=1,N_4} f^{(3,4)}(r_{3i}) \, 
  \prod_{m>l=1,N_4} f^{(4,4)}(r_{lm}) \, , 
\label{eq:F_2}
\end{equation}
and $F_3$ is given by the correspondent product of triplet correlations, 
$f^{(\alpha,\beta,\gamma)}({\bf r}_\alpha,{\bf r}_\beta,{\bf r}_\gamma)$.

The correlation functions are variationally obtained by minimizing 
the g.s. energy of the system, $E_0$. The procedure is outlined in I, 
where a parametrized form for the triplet correlations was used and 
the Jastrow factors were obtained by the Euler equations,  
$\delta E_0/\delta f^{(\alpha \beta)}=0$.
The equations were solved within the HNC framework and the 
{\it scaling} approximation for the elementary diagrams \cite{scaling}. 
The Aziz interatomic potential \cite{Aziz} was used in the minimization 
process.

The perturbative calculation of I included one and two independent 
phonon (OP and TIP, respectively) states and all the diagrams 
corresponding to successive rescatterings of the one phonon states (OPR). 
This contribution was 
obtained by solving a Dyson-like equation in the correlated basis. 
While the correlation factors are intended to care for the short 
range modifications of the ground state wave function due to the 
strongly repulsive interatomic potential, the basic physical effect 
induced by the perturbative corrections may be traced back to the 
backflow around both the impurity and the $^4$He atoms. The CBF analysis 
provided $\mu_3(CBF)=-2.62~K$ (vs. $\mu_3(expt)=-2.79~K$) 
and $m_3^*(CBF)=2.2~m_3$ at equilibrium 
density. The chemical potential was obtained with the Lennard-Jones 
potential and some improvement may be expected by the Aziz interaction.

In order to construct the CBF perturbative series, we 
write $e(q)=e_0(q)+\Delta e(q)$, 
with $e_0(q)=\hbar q^2/2m_3$ and 
\begin{equation}
\Delta e(q)\sim\Delta e_{OP}(q)+\Delta e_{TIP}(q)+\Delta e_{OPR}(q)
 \, . 
\label{eq:Delta}
\end{equation}
The different terms in (\ref{eq:Delta}) represent contributions 
from the corresponding intermediate states.
The $n$-phonon states have been Schmidt-orthogonalized to 
states with a lower number of phonons. For instance, the actual OP 
state reads as:
\begin{equation}
|{\bf q};{\bf q}_1\rangle = \frac {
|\Psi_{{\bf q};{\bf q}_1}\rangle- 
|\Psi_{{\bf q}}\rangle
\langle \Psi_{{\bf q}}|\Psi_{{\bf q};{\bf q}_1}\rangle}{
\langle \Psi_{{\bf q};{\bf q}_1}|\Psi_{{\bf q};{\bf q}_1}\rangle^{1/2}}
 \, . 
\label{eq:OP_0}
\end{equation}
The two-phonon state, $\Psi_{{\bf q};{\bf q}_1{\bf q}_2}$, has been 
orthogonalized in a similar way to $\Psi_{{\bf q}}$, 
$\Psi_{{\bf q};{\bf q}_1+{\bf q}_2}$ and 
$\Psi_{{\bf q};{\bf q}_{1,2}}$. The orthogonalization is a necessary 
step in fastening the convergence of the series as the non orthogonalized 
states have large mutual overlaps 

The non diagonal matrix elements (MEs) of the hamiltonian, $H$, 
(we recall that we use the Aziz potential) are evaluated by 
assuming that the two- and three-body correlations are solutions 
of the corresponding Euler equations. This is not strictly true 
for the triplet correlations but the corrections are expected 
to be small. With this assumption, it is easily verified that
\begin{equation}
\langle{\bf q}|H |{\bf q};{\bf q}_1\rangle = 
-{[N_4 S(q_1)]^{-1/2}}
\frac {\hbar^2}{2 m_3} {\bf q}\cdot {\bf q}_1 S_3(q_1)
 \, , 
\label{eq:OP_1}
\end{equation}
where $S(q_1)$ and $S_3(q_1)$ are the $^4$He and impurity static 
structure functions.

In general, MEs involving $n-1$ phonon states, 
are expressed in terms of the $n$-body structure functions
\begin{equation}
S^{(n)}({\bf q}_1,..{\bf q}_{n})=\frac {1}{N_4} \frac {
\langle \Psi_0 |\rho^\dagger_4({\bf q}_1)..\rho^\dagger_4({\bf q}_{n-1})
\rho_4({\bf q}_n)|\Psi_0\rangle }{\langle\Psi_0|\Psi_0\rangle}
 \, , 
\label{eq:S4_n}
\end{equation}
and
\begin{equation}
S^{(n)}_3({\bf q}_1,..{\bf q}_{n})=\frac {
\langle \Psi_0 |\rho^\dagger_4({\bf q}_1)..\rho^\dagger_4({\bf q}_{n-1})
\rho_3({\bf q}_n)|\Psi_0\rangle }{\langle\Psi_0|\Psi_0\rangle}
 \, , 
\label{eq:S3_n}
\end{equation}
with ${\bf q}_n={\bf q}_1+..+{\bf q}_{n-1}$.

The diagonal MEs have a particularly simple form:
\begin{equation}
\langle {\bf q};{\bf q}_1..{\bf q}_n|
{\bf q};{\bf q}_1..{\bf q}_n\rangle  = 
N_4^n S(q_1)..S(q_n)
 \, , 
\label{eq:diag_1}
\end{equation}
and
\begin{equation}
\langle {\bf q};{\bf q}_1..{\bf q}_n|H|
{\bf q};{\bf q}_1..{\bf q}_n\rangle  = 
E_0^v+e_0(q)+\sum_{i=1,n}w_F(q_i)
\label{eq:diag_2}
\end{equation}
with $E_0^v=\langle \Psi_0|H|\Psi_0\rangle/\langle \Psi_0|\Psi_0\rangle$ and 
$w_F(q_i)=\hbar^2 q_i^2/2m_4S_4(q_i)$ is the Feynman $^4$He 
excitation spectrum\cite{Feynman}.

The OP and TIP perturbative diagrams contributing to $\Delta e(q)$ are
shown in Fig.(5) of I, where only their $q=0$ derivative was 
computed, since the paper was concerned with just the calculation 
of the effective mass at $q=0$. Here we extend the formalism to 
finite $q$. We use Brillouin-Wigner perturbation theory, so 
the correction itself depends on $e(q)$ and the series must be 
summed self-consistently. For instance, the OP contribution 
is given by
%\begin{equation}
\begin{eqnarray}
\nonumber
\Delta e_{OP}(q)&=&\sum_{{\bf q}_1} \frac {
\vert \langle {\bf q}|H-E_0-e(q)|{\bf q};{\bf q}_1\rangle \vert ^2
}{
e(q)-e_0(|{\bf q}-{\bf q}_1|)-w_F(q_1)}
 \\  & = &
\frac {\Omega}{(2\pi)^3}
\left(\frac {\hbar^2}{2m_3}\right)^2
\int d^3q_1
\frac {1}{N_4S(q_1)}
\frac {[S_3(q_1){\bf q}\cdot{\bf q}_1]^2}{
e(q)-e_0(|{\bf q}-{\bf q}_1|)-w_F(q_1)}
 \, . 
\label{eq:OP_3}
\end{eqnarray}
%\end{equation}

The expressions of the other diagrams are quite lengthy and will 
not be reported here. However, some comments are in order. They 
involve the two- and three-body structure functions, {\it i.e.} 
the Fourier transforms of the two- and three-body 
distribution functions, $g^{(2)}(r_{12})$ and 
$g^{(3)}({\bf r}_1,{\bf r}_2,{\bf r}_3)$. $g^{(2)}$ is a direct 
output of the HNC/Euler theory and, in pure $^4$He, results to 
be very close to its experimental measure. To evaluate $g^{(3)}$ 
is more involved and usually one has to resort to some approximations.
The mostly common used are the convolution approximation (CA) and 
the Kirkwood superposition approximation (KSA)\cite{Feenberg}. 
The CA correctly accounts for the sequential relation between 
$g^{(3)}$ and $g^{(2)}$ and factorizes in momentum space, 
$S^{(3)}_{CA}({\bf q}_1,{\bf q}_2,{\bf q}_3)=S(q_1)S(q_2)S(q_3)$; 
the SA factorizes in $r$-space, 
$g^{(3)}_{KSA}({\bf r}_1,{\bf r}_2,{\bf r}_3)=
g^{(2)}(r_{12})g^{(2)}(r_{13})g^{(2)}(r_{23})$, and describes 
adequately the short range region. The momentum space factorization 
property makes the use of the CA particularly suitable for 
our perturbative study.

The sensitivity of the calculation to the approximation for $g^{(3)}$ 
clearly shows up in the CBF-TIP evaluation of the $^4$He excitation 
spectrum, $\omega(q)$. Figure 1 compares the Feynman spectrum and those 
obtained within the CA and KSA with the experimental data. The phonon 
linear dispersion at low-$q$ is well reproduced by both $\omega_F(q)$ and 
$\omega_{CA}(q)$, whereas $\omega_{KSA}(q)$ fails to give the 
correct behavior. As it is well known, the remaining part of the 
spectrum is severely overestimated by $\omega_F(q)$; both CA and KSA 
give a reasonable description of the maxon region but KSA is closer 
to the experiments at the roton minimum, just in view of its better 
description of the short-range regime. An overall good agreement with 
the experimental curve was obtained in ref.\cite{Manousakis86} where 
backflow correlations were added to the CBF states.

Figure 2 shows $e(q)$ in CA and KSA, along with the data from 
ref.\cite{Fak90}. The curves do not include the OPR  
contribution. At this level, the effective masses are 
$m^*_3(CA)=1.6~m_3$ and $m^*_3(KSA)=2.1~m_3$ and, again, KSA 
is closer to the experimental spectrum at large momenta. The 
curve labelled as CA1 is obtained in CA, but using the experimental 
$^4$He spectrum in the energy denominators. Diagram (5.e) of I, 
that gives the two-phonon correction to $\omega(q)$, has not been 
included as its effect is mostly taken into account by the use of 
$\omega_{expt}(q)$. KSA and CA1 are close at large $q$-values, 
pointing to a good  description of the $^4$He roton as a key ingredient 
for a correct approach to the large $q$ sector. We will follow 
the CA1 method for the remaining of the work.

Figure 3 gives the CA1 impurity spectrum and the experimental 
$^3$He and $^4$He curves. The OPR terms are included and 
the LP and LPM fits to $e_{expt}(q)$ are shown. Since the 
branch of the dynamical response due to the excitations of the low 
concentration $^3$He component in the Helium mixtures overlaps 
the collective $^4$He excitation at $q>1.7~\AA^{-1}$ 
\cite{Fak90,Fabrocini97}, $e_{expt}(q)$ is not known in 
that region. A roton-like behavior was supposed in ref.\cite{Pita73}. 
This structure was not confirmed by the variational MonteCarlo (VMC) 
calculation of ref.\cite{Reatto96}, which employed shadow wave functions 
in conjuction with a Jastrow correlation factor of the McMillan 
type. The VMC data at equilibrium density are given in the 
figure: they overestimate the experiment and have an effective 
mass of $m^*_3(VMC)\sim 1.7~m_3$.  

The shadow wave function of ref.\cite{Reatto96} takes into 
account backflow effects. Actually, in several papers was 
pointed out that second order perturbative expansion with 
OP states introduces backflow correlations into the wave 
function \cite{Fabrocini86,Davison69,McMillan69}. We find 
$m^*_3(OP)=1.8~m_3$, in good agreement 
with the VMC outcome. An analogous CBF treatment by 
Saarela \cite{Saarela90} gave similar results ($m_3^*\sim 1.9~m_3$) 
and a spectrum close to the LP form. More complicated momentum dependent 
correlations are generated by TIP and OPR diagrams, 
playing  a relevant role in the CBF approach and 
giving $m^*_3(CBF)=2.1~m_3$. 

The total CBF impurity spectrum is very close to $e_{expt}(q)$ 
up to its merging into the $^4$He dispersion relation. 
For the $\gamma$  parameter in the LPM parametrization, the theory gives 
$\gamma (CBF)\sim 0.19~\AA^2$. If the spectrum is parametrized 
in terms of a momentum dependent effective mass, 
$e(q)=\hbar^2 q^2/2m_3^*(q)$, then we find 
$m^*_3(q=1.7~\AA^{-1})=3.2~m_3$, with an increase of $\sim 50\%$ 
respect to the $q=0$ value. 
 
Beyond $q\sim 1.9~\AA^{-1}$, the energy denominators vanish for 
some momentum values and the series cannot be summed anymore. This 
is due to the fact that the impurity quasiparticle 
is no longer an excitation with a well defined 
energy, since it can decay into $^4$He excitations and acquire a 
finite lifetime, $\tau$. A finite $\tau$-value reflects a non zero 
imaginary part of the $^3$He complex optical potential (or the 
on-shell self-energy), $W(q)=\Im \Sigma (q,e(q))$\cite{Fantoni83}. 
Figure 3 shows $W(q)$ as computed with only OP intermediate states,
\begin{equation}
W_{OP}(q)=\pi\sum_{{\bf q}_1} 
\vert \langle {\bf q}|H-E_0-e(q)|{\bf q};{\bf q}_1\rangle \vert ^2
\delta(e(q)-e_0(|{\bf q}-{\bf q}_1|)-w(q_1))
 \, , 
\label{eq:opt}
\end{equation}
where the LPM impurity spectrum and the experimental $^4$He 
dispersion have been used (notice that $W(q)$ is amplified by 
a factor 4 in the figure). The OP optical potential is close to the 
one found in ref.\cite{Saarela90}. A numerical extrapolation of the 
computed $e_{CBF}(q)$ into the roton region does not show any 
evidences of a $^3$He roton-like structure. 

In conclusion, we find that CBF perturbative theory is able 
to give a quantitative description of the $^3$He impurity 
excitation spectrum in liquid $^4$He at equilibrium density. 
The intermediate correlated states must consider at least 
two independent phonon states and one phonon state rescattering 
is found to play a non marginal role at large momenta. 
It is plausible that in a richer basis, including for instance 
explicit backflow correlations, a lower order expansion might 
be sufficient. However, the more complicated structure of the 
matrix elements could cause additional uncertainties in their 
evaluation, at least in the framework of the cluster expansion 
approach. The development of a MonteCarlo based alghoritm for 
the computation of non diagonal matrix elements would probably be 
the correct answer. 

% I am here

{\bf ACKNOWLEDGEMENTS}

A.F. wants to thank 
the Institute for Nuclear Theory at the University of Washington for 
its hospitality and the Department of Energy for its partial support 
during the completion of this work.
This research was also partially supported by DGICYT (Spain) Grant
No. PB95-1249, the agreement CICYT (Spain)--INFN (Italy) and 
the Acci\'on Integrada program (Spain-Italy).

\begin{figure}
\caption{ $^4$He excitation spectrum at equilibrium density.
Stars are the experimental data. Energies in K and momenta in $\AA^{-1}$.}
\label{fig:1}
\end{figure}
 
\begin{figure}
\caption{ $^3$He single particle energies in CA, KSA and CA1 without 
phonon rescattering. Stars are the experimental data. Units as in fig.(1).}
\label{fig:2}
\end{figure}
 
\begin{figure}
\caption{ CBF/CA1 (triangles), LP and LPM $^3$He single particle energies. 
Full circles are the VMC data. Stars and circles  are 
the impurity and $^4$He experimental data, respectively. Black triangles 
are extrapolated CBF/CA1 values (see text). Black diamonds give the 
impurity imaginary optical potential (in K). Units as in fig.(1).}
\label{fig:3}
\end{figure}
 
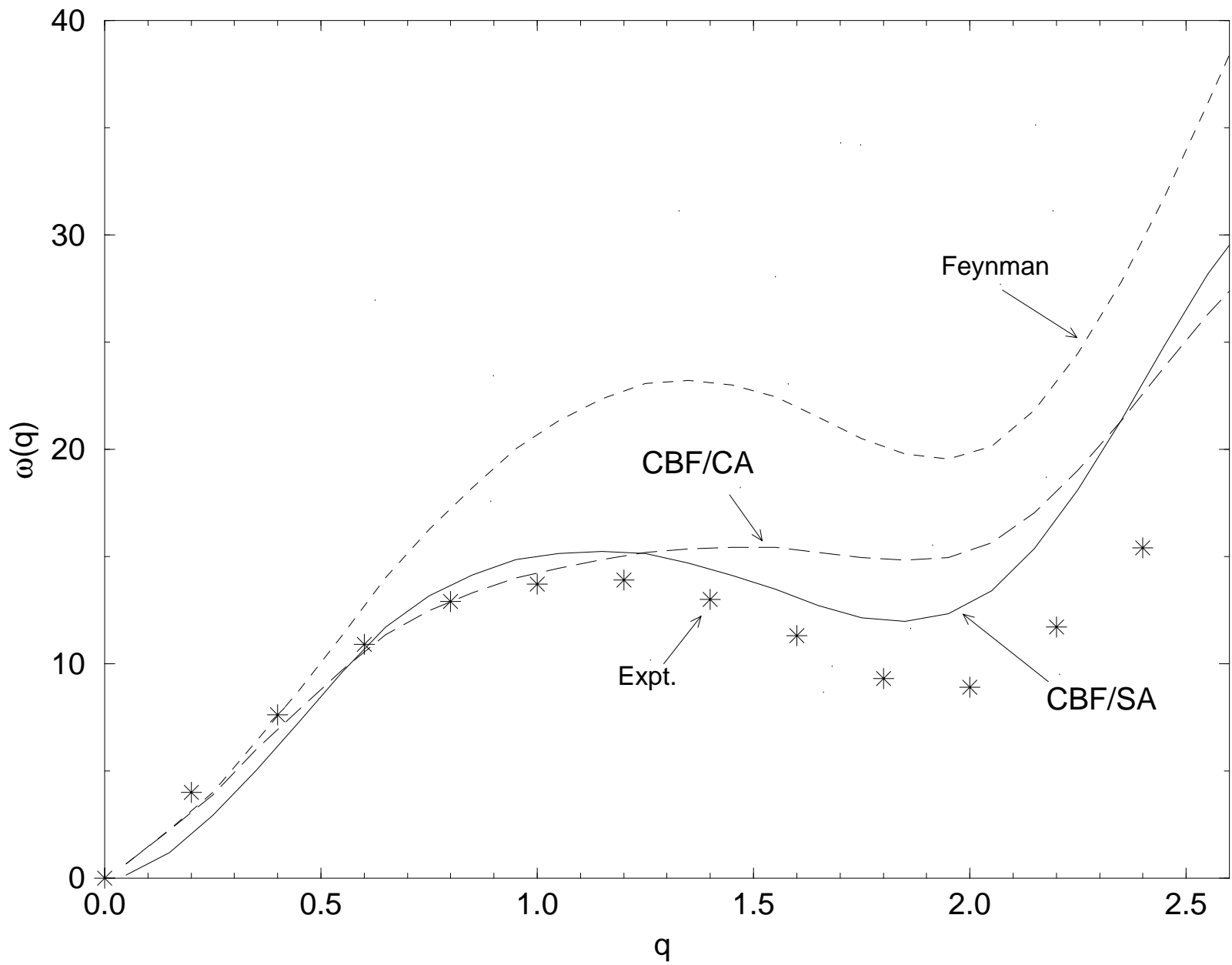
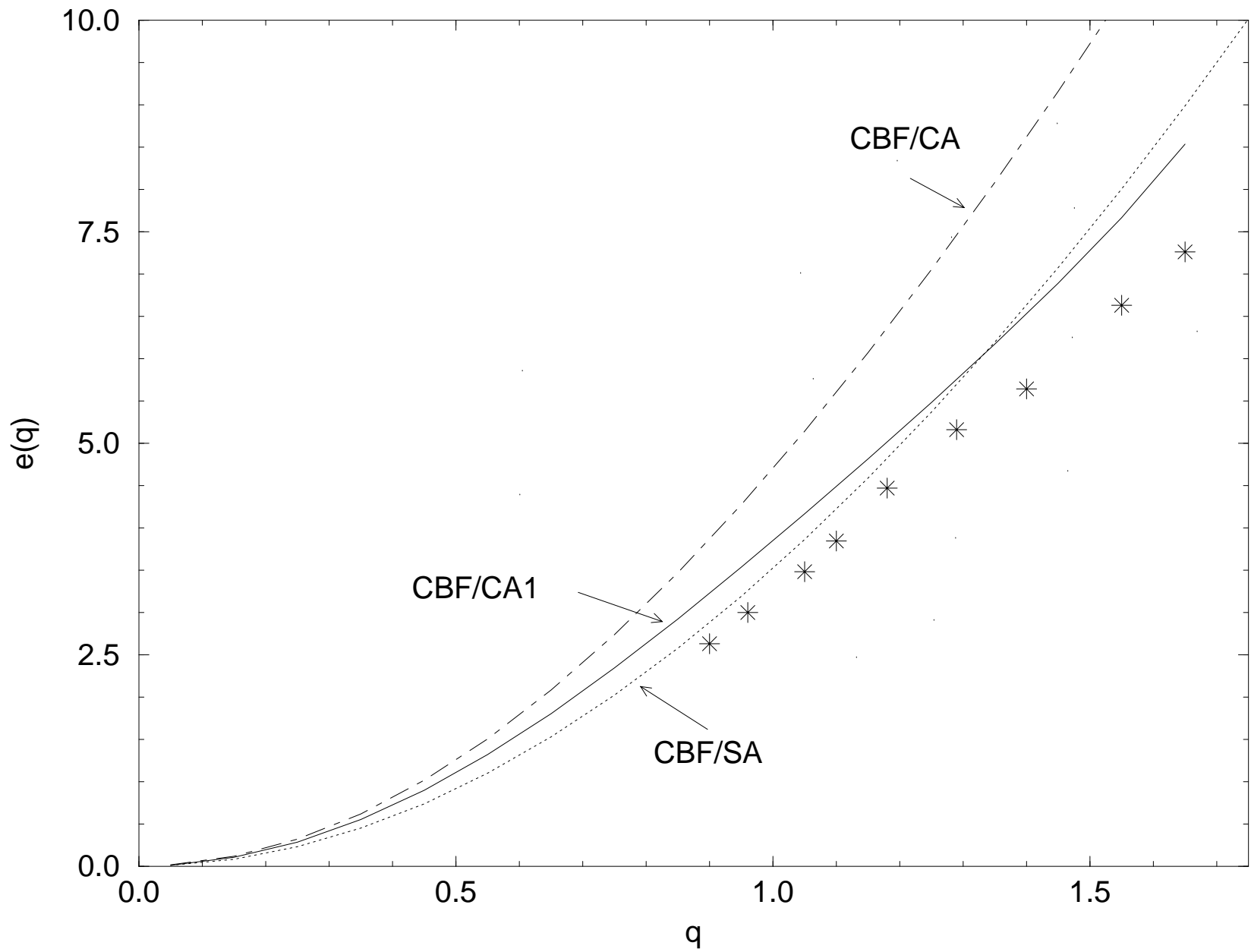
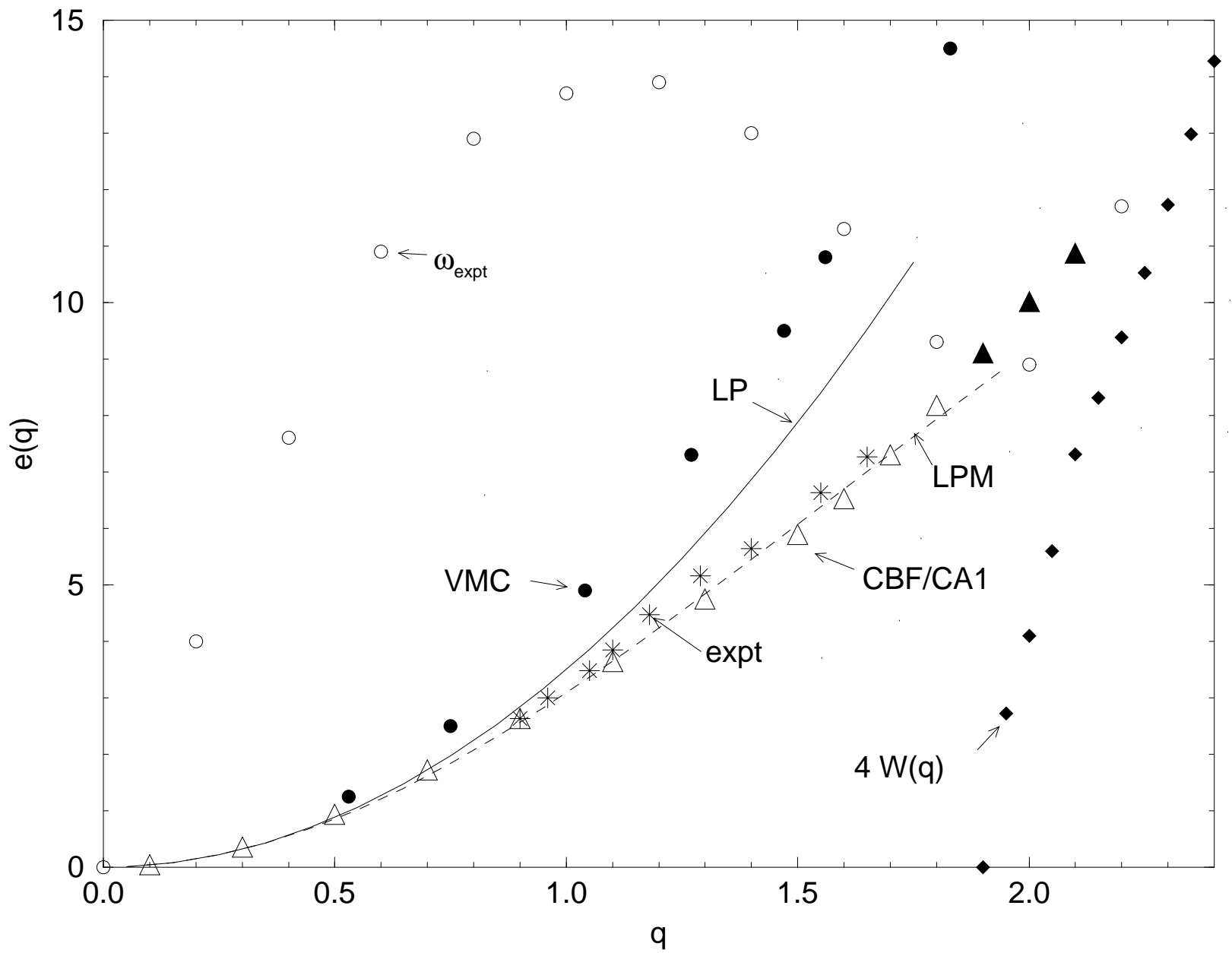
\end{document}